\input{aipcheck}

\documentclass[
    ,final            
  ]
  {aipproc}

\layoutstyle{6x9}

\begin{document}

\title{Recent Type II Radio Supernovae}

\classification{97.60.Bw, 95.85.Bh, Fm}
\keywords      {Radio Supernovae; SN~1993J, SN~2001gd, SN~2001em, SN~2002hh, SN~2004dj, SN~2004et}

\author{C. J. Stockdale}{
  address={Physics Dept., Marquette University, PO Box 1881, Milwaukee, WI 53201},
 email={christopher.stockdale@mu.edu},
}

\author{M. T. Kelley}{
  address={Physics Dept., Marquette University, PO Box 1881, Milwaukee, WI 53201}
}
\author{K. W. Weiler}{
  address={Naval Research Laboratory}
}
\author{N. Panagia}{
  address={Space Telescope Science Institute}
}
\author{R. A. Sramek}{
  address={National Radio Astronomy Observatory}
}
\author{J. M. Marcaide}{
  address={University of Valencia}
}
\author{C. L. M. Williams}{
  address={Massachusetts Institute of Technology}
}
\author{S. D. Van Dyk}{
  address={Spitzer Science Center, California Institute of Technology}
}

\begin{abstract}
We present the results of radio observations, taken primarily with the Very Large Array, of Supernovae~1993J, 2001gd, 2001em, 2002hh, 2004dj, and 2004et.  We have fit a parameterized model to the multi-frequency observations of each supernova.  We compare the observed and derived radio properties of these supernovae by optical classification and discuss the implications. 
\end{abstract}

\maketitle


\section{Introduction}

We present the results of six years of radio observations of type II supernovae (SNe) using the The Very Large Array (VLA)\footnote{The VLA telescope of the National Radio Astronomy Observatory is operated by the Associated Universities, Inc. under a cooperative agreement with the National Science Foundation.}.  The radio emission from core-collapse SNe results from the interaction between the expanding SN blastwave ($v_{\rm blast} \sim 10^4 {\rm km} {\rm s^{-1}}$) and the circumstellar material (CSM) shed by the progenitor star in the $\sim 10,000$ years preceding the SN explosion ($v_{\rm wind} \sim 10 {\rm km} {\rm s^{-1}}$.  The electrons in the CSM were photo ionized by the SN explosion and have a temperature $\sim 10^4$ K.  The model for radio emission was explained by Chevalier (1982a, 1982b) and our parameterized model which we fit our data is explained in Sramek et al. (2005) and references therein.  A cartoon version of this model is presented in Figure 1.  Here present the radio observations of SN~2001gd (type IIb; NGC~5033), SN~2001em (type IIn; UGC~11794), SN~2002hh (type II; NGC~6946), SN~2004dj (type IIP; NGC~2403), and SN~2004et (type IIP; NGC~6946).  The radio light curves for these SNe are presented in Kelley et al. (2007).  We compare the measured properties of these recent SNe with the 14 years of radio observations of SN ~1993J and other historical SNe (Weiler et al. 2007; Stockdale et al. 2006). 

\begin{figure}
 \rotatebox{270}{\includegraphics[height=.3\textheight]{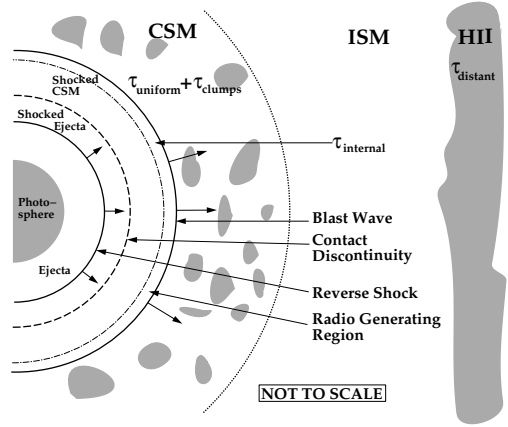}}
  \caption{Not to scale model identifying the radio emitting region associated with SN explosions and the various absorbing regions.}
\end{figure}

\section{Model Description}

Figure 1 identifies the regions significant to production and absorption of radio emission from the expanding blastwave of a core-collapse supernova.  Table 1 indicates the derived fits to our observations of the recent radio supernovae.  The parameters listed in Table 1 correspond to the flux density ($K_1$), homogeneous ($K_2$, $K_4$), and clumpy or filamentary ($K_3$) free-free absorption (FFA) at 5~GHz, formally but not necessarily physically, one day after the explosion date $t_0$.  The $K_2$ absorption terms represents local, homogeneous free-free absorbing CSM and clumpy or filamentary free-free absorbing CSM, respectively, that are near enough to the SN progenitor that they are altered by the rapidly expanding SN blastwave.  The $K_4$ absorption term represents FFA that is produced by an ionized medium that completely covers the emitting source (``homogeneous external absorption'') and is consequently constant in time.  None of our observations indicate the necessity of including a $K_4$ term. The $K_3$ term describes the attenuation produced by an inhomogeneous FFA medium.  All external and clumpy absorbing media are assumed to be purely thermal, singly ionized gas which absorbs via FFA with frequency dependence $\nu^{-2.1}$ in the radio.  

The FFA optical depth outside the emitting region is proportional to the
integral of the square of the CSM density over the radius.  Since in the
simple Chevalier (1982a, 1982b) model the CSM density decreases as $r^{-2}$, the
external optical depth will be proportional to $r^{-3}$.

Since it is physically realistic and may be needed in some RSNe where radio observations have been obtained at early times and high frequencies, our parameterized model includes the possibility for an internal absorption term.  This internal absorption term may consist of two parts -- synchrotron self-absorption (SSA) which is represented by the $K_5$ term, and mixed, thermal FFA/non-thermal emission which represented by the $K_6$ term.   Our data indicates that some of the radio SNe require a $K_5$ component but none indicate the necessity of $K_6$ component.   $K_5$ corresponds to the internal, non-thermal ($\nu^{\alpha - 2.5}$) SSA and $K_6$ corresponds to the internal thermal ($\nu^{-2.1}$) free-free absorption mixed with non-thermal emission, at 5~GHz, formally but not necessarily physically, one day after the explosion date $t_0$. 

The various $\delta$ parameters describe the time dependence of the optical depths for their respective time-varying absorption terms.   The unabsorbed spectral indices ($\alpha$; $S_{\nu}\propto \nu^{+\alpha}$) and the unabsorbed time decay parameters ($\beta$; $S_{\nu}\propto t^{+\beta}$) are also presented in Table 1.

\begin{table}
\tiny{
\begin{tabular}{ccccccc}
\hline
    \tablehead{1}{c}{b}{Parameter}
  & \tablehead{1}{c}{b}{SN~1993J}
  & \tablehead{1}{c}{b}{SN~20001gd}
  & \tablehead{1}{c}{b}{SN~2001em}
  & \tablehead{1}{c}{b}{SN~2002hh}
  & \tablehead{1}{c}{b}{SN~2004dj}
  & \tablehead{1}{c}{b}{SN~2004et} \\
\hline
$K_1$ & $4.6 \times 10^3$ & $7.5\times 10^2$ & $8.0 \times 10^1$ &  $2.6 \times 10^1$ & $1.5\times 10^1$ & $4.3 \times 10^1$\\
$\alpha$ & $-0.81$ & $-0.94$ & $-0.70$ & $-1.10$ & $-1.08$ & $-1.05$   \\
$\beta$ & $-0.73$ & $-0.92$ & $-0.49$ & $-0.58$ & $-0.69$ & $-0.60$    \\ 
$K_2$ & $1.6 \times 10^2$ & $1.5 \times 10^3$ & $\equiv 0.00$ & $\equiv 0.00$ &  $1.4 \times 10^0$ & $\equiv 0.00$ \\
$\delta$ & $-1.88$ & $-1.88$   &   &   & $-1.00$ &   \\
$K_3$ & $4.6 \times 10^5$ & $2.1 \times 10^8$ & $4.1 \times 10^9$ & $3.0 \times 10^3$ & $9.7 \times 10^0$ & $5.1 \times 10^0$ \\
$\delta^{\prime}$  & $-2.83$ & $-3.6$ & $-3.24$ & $-2.05$  & $-1.14$ & $-2.16$\\ 
$K_5$ & $2.6 \times 10^3$ & $1.7 \times 10^2$ & $\equiv 0.00$ & $\equiv 0.00$ & $\equiv 0.00$ & $1.2 \times 10^0$ \\
$\delta^{\prime\prime}$  & $-2.05$  & $-1.5$  &  &  &  & $-1.43$\\ 
Time to $L_{\rm 6\ cm\ peak}$ \ (days) & 133 & 80 & 1320 & 62.2 & 7.12 & 66.3 \\
$L_{\rm 6\ cm\ peak}$\ (erg s$^{-1}$ Hz$^{-1}$) & $4.8 \times 10^{27}$ &  $3.8 \times 10^{27}$ & $1.9 \times 10^{28}$ & $8.5 \times 10^{25}$ &  $4.2 \times 10^{25}$ & $1.2 \times 10^{26}$ \\
${\rm \dot M}$ (${\rm M_\odot}$ yr$^{-1}$) & & $1.0 \times 10^{-5}$ & $2.5 \times 10^{-4}$ & $3.6 \times 10^{-6}$ & $8.2 \times 10^{-6}$ & $1.2 \times 10^{-7}$  \\
\hline
\end{tabular}}
\caption{Derived parameters for the model described in Sramek et al. (2005).  Assumes a constant mass-loss rate, constant velocity, wind-established CSM, i.e., $\rho \propto r^{-2}$.}
\label{tab:a}
\end{table}

\section{Results of Recent Observations of SN~2001gd}

An extensive search for the best parameter fits to our recent supernovae observations were made and are presented in Table 1.  The data are presented in radio light curves in Kelley et al. (2007).  Our results indicate that the CSM of most of these supernovae are best described as purely filamentary,  or clumpy with no clear evidence of SSA present in the radio light curves.  SN~1993J is discussed at length in Weiler et al. (2007) and we include it here for comparison to the recent SN observations.  Of the recent radio SNe, only SN~2004et is well sampled enough to require a SSA component in the parameterized model, as the other SNe lack sufficient early data to distinguish the initial nature of the absorption mechanism.  Since data is still being gathered on SNe~2001em, 2002hh, 2004dj, and 2004et, we will limit the remaining discussion to SN~2001gd.

However, we have estimated the brightness temperatures for the ``early'' 20~cm emission based on a pure FFA model and found that they exceed the limiting temperature for synchrotron emission of $\sim10^{11.5}$ K calculated by Readhead (1994), which was also the case for SN~1993J (Weiler et al. 2007).  Given our inability to model this early absorption based purely on our sparse ``early'' data, we have chosen to assume that SN~2001gd has a similar brightness temperature evolution ($\sim10^{11}$ K at peak) for the 20~cm emission as it becomes optically thin. Given that SN~2001gd and SN~1993J have similar characteristics at ``early'' times, we estimate, as was the case for SN~1993J, a significant SSA component is required for SN~2001gd along with the FFA component determined from fitting which describes the rapid rise of the 20~cm data well. The best parameters which we can estimate for the ``early'' epoch of SN~2001gd with these assumptions are given in Table~1. We note an apparent similarity between the $\delta$ and $\delta^{\prime\prime}$ terms from our modeling of the SN~2001gd radio emission which is similar to the case of SN~1993J where the two values were $\delta=-1.88$ and $\delta^{\prime\prime} = -2.05$, respectively.  However, it must be stated that, from pure modeling, our sparse ``early'' data only constrain the values of  $\delta$ and $\delta^{\prime\prime}$ for SN~2001gd to lie between $-1$ and $-2$. To narrow the range of these values, we have therefore assumed a value of $-1.88$ for $\delta$, to match the value for $\delta$ determined for SN~1993J (Weiler et al. 2007).

The ``late'' period appears to be optically thin at all observed frequencies with a much steeper decline rate than is possible with the ``early'' data determined $\beta = -0.92$. Again, if we assume a similarity to the exponential decline seen for SN~1993J \citep{Weiler07}, the late epoch of SN~2001gd is consistent with an exponential decline after day 550, with an e-folding time of 500 days.  This observed ``break" in the radio light curves of SN~2001gd is interpreted as indicative of a rapid transition of the blastwave into a much more tenuous medium.  This occurs much earlier for SN~2001gd (day $\sim550$) than that seen for SN~1993J (day $\sim3,100$) by Weiler et al. (2007).

Before the break, SN~2001gd is still somewhat optically thick at 20~cm, but after the break it appears to be optically thin.  This discounts the possibility of the break being the result of a sudden change in the shock velocity or magnetic field properties at the shock front since such changes would not yield an optical depth reduction. Given the similar expansion velocities of $\leq 13,000$ km s$^{-1}$ for SN~2001gd and $11,000 - 15,000$ km s$^{-1}$ for SN~1993J (P{\'e}rez-Torres et al. 2005; Marcaide et al. 2007), it is clear that both supernovae had a period of enhanced mass-loss rate preceding their explosions.  Assuming a ``standard'' blastwave velocity for SN~2001gd of $10,000$ km s$^{-1}$  and a ``standard'' pre-explosion wind velocity of 10 km s$^{-1}$, this enhanced mass-loss persisted for the $\sim$1,500 years prior to the explosion.  However in the case of SN~1993J, this phase began $\sim$ 8,800 years earlier than the explosion of its progenitor star. The shorter duration of the high pre-supernova mass-loss rate for SN~2001gd is also born out in its faster rise and brighter luminosity, indicating there was less intervening CSM to contribute to the FFA at early epochs. 

\section{Preliminary Results from Recent Radio Supernovae}
From our study of these six supernovae, we are able to make the following statements.
\begin{itemize}

\item The spectral indices appear to steepen from $-0.70$ for IIn SNe to $-1.10$ for IIP SNe.
\item The time decay parameters for IIn SNe is small $-0.50$ for IIn SNe as compared to the limiting value of $-0.75$ measured for IIP and IIb SNe.
\item The peak luminosities (erg s$^{-1}$) increase from $10^{25}$ (IIP) to $10^{27}$ (IIb) to $10^{28}$ (IIn). 
\item The estimated mass loss rates (${\rm M_\odot}$ yr$^{-1}$) follow a similar trend of $10^{-6}$ (IIP) to $10^{-5}$ (IIb) to $10^{-4}$ (IIn).
\end{itemize}

Our results indicate that Type IIn SNe have significant absorption due to the CSM ejected by the progenitor star, typically becoming optically thin years after the explosion date.  Type IIb SNe, have much less CSM and it tends to be clumpy in nature being ejected over a period of a few thousand years prior to the progenitor explosion ({\it e.g.} SNe~1993J and 2001gd).
Finally, Type IIP SNe appear to have very little CSM resulting in weak radio emission that rises and fades quickly with time.  A long-term goal of our program to observe radio emission from SNe is to chart the evolution of SNe from the explosion of the progenitor star into a supernova remnant.  Figure 2 illustrates the 20 cm radio light curves of a number of historical supernovae, with the Cas A and Crab remnants identified for comparison (Stockdale et al. 2006).

\begin{figure}
  \includegraphics[height=.3\textheight]{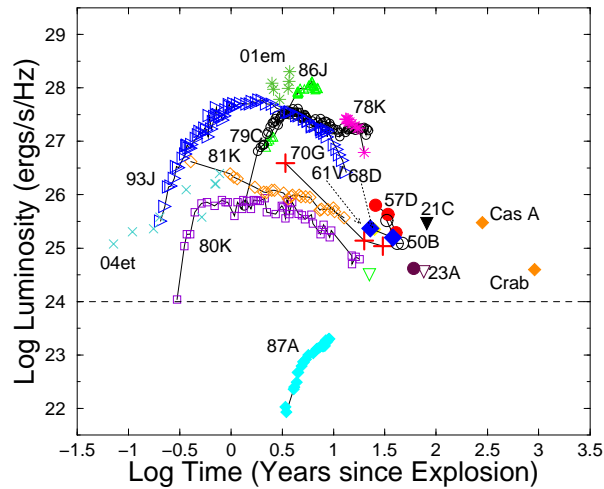}
  \caption{20 cm radio light curves for historical SNe, see Stockdale et al. (2006) for complete references.}
\end{figure}

\begin{theacknowledgments}
CJS is a Cottrell Scholar, supported by Research Corporation.  MTK was supported by the NASA Wisconsin Space Grant Consortium.  KWW thanks the Office of Naval Research for the 6.1 funding which supports his research.
\end{theacknowledgments}


\begin{thebibliography}{9}
\bibitem{Chevalier82a} 
Chevalier, R.\ A. 1982a, ApJ, 259, 302
\bibitem{Chevalier82b} 
Chevalier, R.\ A. 1982b, ApJL, 259, L85
\bibitem{Kelley07}
Kelley, M.T. et al. "Light Curves of Radio Supernovae" in \emph{SN~1987A: 20 Years After - Supernovae and Gamma-Ray Bursters} ed. by Weiler, K. W., and Immler, S., AIP, U.S.A., 2007.
\bibitem{Marcaide07} 
Marcaide, J.\ M. et al. 2007, MNRAS, submitted
\bibitem{PerezTorres05} 
P{\'e}rez-Torres, M.~A., et al.\ 2005, MNRAS, 360, 1055 
\bibitem{Readhead94} 
Readhead, A.~C.~S. 1994, ApJ, 426, 51
\bibitem{Sramek05}
Sramek et al. "Radio Supernovae" in \emph{Cosmic Explosions On the 10th Anniversary of SN 1993J}, ed. by Marcaide, J.M., and Weiler, K.W., Springer-Verlag Berlin Heidelberg, Germany, 2005
\bibitem{Stockdale07}
Stockdale, C. J. et al. 2006, AJ, 131, 2, 889
\bibitem{Weiler07} 
Weiler, K.\ W. et al. 2007, ApJ, in press
\end{thebibliography}
\end{document}